\documentclass[prl,superscriptaddress,twocolumn,showpacs,reprint]{revtex4-1}
\usepackage{color,graphicx,calc,url}
\usepackage{dcolumn}
\usepackage{bm}
\usepackage{amssymb}
\usepackage{amsmath}
\usepackage{wasysym}
\usepackage{hyperref}
\usepackage{siunitx}
\usepackage{mhchem}
\usepackage{tabularx}
\newcolumntype{L}[1]{>{\raggedright\arraybackslash}p{#1}} 
\newcolumntype{C}[1]{>{\centering\arraybackslash}p{#1}} 
\newcolumntype{R}[1]{>{\raggedleft\arraybackslash}p{#1}} 

\begin{document}

\title{Longitudinal spin Seebeck effect contribution in transverse spin Seebeck effect experiments in Pt/YIG and Pt/NFO}

\author{D.\ Meier}
\email{dmeier@physik.uni-bielefeld.de}
\homepage{www.spinelectronics.de}
\author{D.\ Reinhardt}
\author{M.\ van Straaten}
\author{C.\ Klewe}
\affiliation{Department of Physics, Center for Spinelectronic Materials and Devices, Bielefeld University, 33501 Bielefeld, Germany}
\author{M.\ Althammer}
\affiliation{Walther-Meissner-Institut, Bayerische Akademie der Wissenschaften, Walther-Meissner-Strasse 8, 85748 Garching, Germany}
\author{M.\ Schreier}
\affiliation{Walther-Meissner-Institut, Bayerische Akademie der Wissenschaften, Walther-Meissner-Strasse 8, 85748 Garching, Germany}
\affiliation{Physik-Department, Technische Universit\"at M\"unchen, 85748 Garching, Germany}
\author{S.T.B.\ Goennenwein}
\affiliation{Walther-Meissner-Institut, Bayerische Akademie der Wissenschaften, Walther-Meissner-Strasse 8, 85748 Garching, Germany}
\affiliation{Nanosystems Initiative Munich (NIM), Schellingstra{\ss}e 4, 80799 M\"unchen, Germany}
\author{A.\ Gupta}
\affiliation{Center for Materials for Information Technology, University of Alabama, Tuscaloosa Alabama 35487, USA}
\author{M.\ Schmid}
\author{C.H.\ Back}
\affiliation{Institute of Experimental and Applied Physics, University of Regensburg, 93040, Germany}
\author{J.-M.\ Schmalhorst}
\author{T.\ Kuschel}
\author{G.\ Reiss}
\affiliation{Department of Physics, Center for Spinelectronic Materials and Devices, Bielefeld University, 33501 Bielefeld, Germany}

\date{\today}


\begin{abstract}

We investigate the inverse spin Hall voltage of a \SI{10}{\nano\meter} thin Pt strip deposited on the magnetic insulators \ce{Y3Fe5O12} (YIG) and \ce{NiFe2O4} (NFO) with a temperature gradient in the film plane. We observe characteristics typical of the spin Seebeck effect, although we do not observe a change of sign of the voltage at the Pt strip when it is moved from hot to cold side, which is believed to be the most striking feature of the \textit{transverse} spin Seebeck effect. Therefore, we relate the observed voltages to the \textit{longitudinal} spin Seebeck effect generated by a parasitic out-of-plane temperature gradient, which can be simulated by contact tips of different material and heat conductivities and by tip heating. This work gives new insights into the interpretation of transverse spin Seebeck effect experiments, which are still under discussion.

\end{abstract}

\maketitle

Spin caloritronics is an active branch in spintronics \cite{Bauer:2012fq,Wolf:2001fu}. The interplay between heat, charge and spin transport opens a new area of fascinating issues involving the use of waste heat in electronic devices. One potentially useful effect for heat harvesting \cite{Kirihara:2012jq} is the spin Seebeck effect (SSE) \cite{Uchida:2008cc} which was observed in 2008.

\begin{figure}[!b]
\centering
\includegraphics[width=\linewidth]{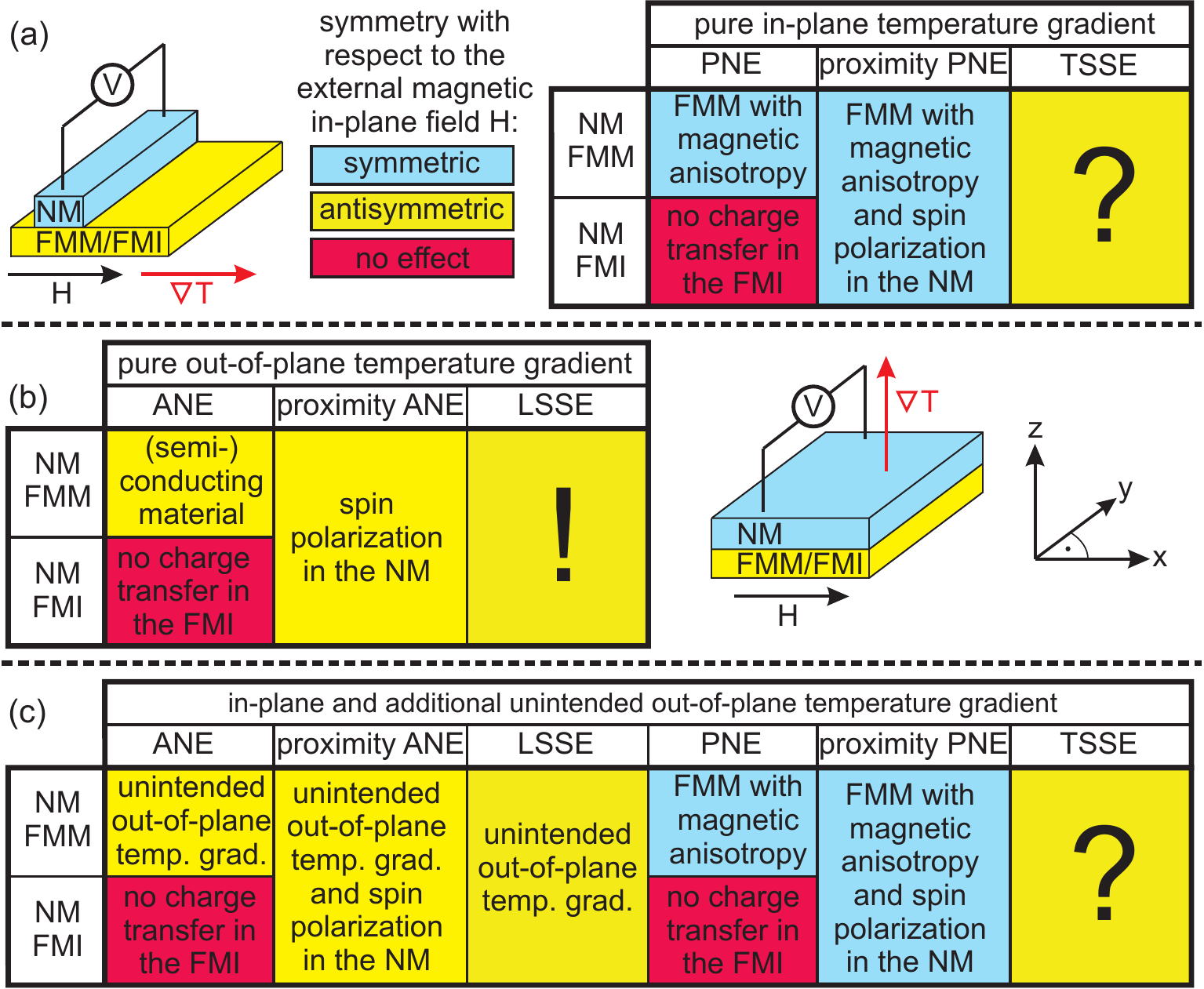}
\caption{Summary of effects in SSE experiments for NM/FMM and NM/FMI bilayers and given symmetry with respect to the external magnetic in-plane field \(H\) for antisymmetric magnetization reversal processes. (a) Pure in-plane \(\nabla T\) for TSSE measurements. (b) Pure out-of-plane \(\nabla T\) for LSSE measurements. (c) In-plane and unintended out-of-plane \(\nabla T\).}
\label{fig:parasiticeffects}
\end{figure}

It was reported that a spin current perpendicular to an applied temperature gradient can be generated in a ferromagnetic metal (FMM) by the transverse spin Seebeck effect (TSSE) \cite{Uchida:2008cc}. An adjacent normal metal (NM) converts the spin current via the inverse spin Hall effect (ISHE) \cite{Saitoh:2006kk} into a transverse voltage, which is antisymmetric with respect to the external magnetic field \(H\) (\(V(H)\!=\!-V(-H)\), cf. Fig.~\ref{fig:parasiticeffects}~(a)). In this geometry, the temperature gradient is typically aligned in-plane (\(\nabla T_\text{x}\)) and can also induce a planar Nernst effect (PNE) in FMM with magnetic anisotropy \cite{Avery:2012bj} which is due to the anisotropic magnetic thermopower and symmetric with respect to \(H\) (\(V(H)\!=\!V(-H)\)). For pure \(\nabla T_\text{x}\) in a ferro(i)magnetic insulator (FMI) there is no PNE, since there are no free charge carriers available. However, if the NM material is close to the Stoner criterion, a static magnetic proximity effect could induce a so called proximity PNE, which in general is present in spin polarized NM adjacent to a FMM and could also occur in a NM-FMI contact (cf. Fig.~\ref{fig:parasiticeffects}~(a)).

For the longitudinal SSE (LSSE) \cite{Uchida:2010jb} the spin current flows directly from the FM into an adjacent NM parallel to the temperature gradient (cf. Fig.~\ref{fig:parasiticeffects}~(b)), which is typically aligned out-of-plane (\(\nabla T_\text{z}\)). In NM/FMM bilayers the anomalous Nernst effect (ANE) can occur, but is absent in the FMI. In semiconducting materials the ANE contributes to the LSSE as already shown for Pt/NFO at room temperature \cite{Meier:2013dz}. Additionally, if the NM would be spin polarized by the proximity to the FM, an additional proximity ANE could occur \cite{Guo:2014bz} (cf. Fig.~\ref{fig:parasiticeffects}~(b)).

We would like to point out that Figs.~\ref{fig:parasiticeffects}~(a) and \ref{fig:parasiticeffects}~(b) include results of previous experiments with pure \(\nabla T_\text{x}\) on NM/FMM \cite{Uchida:2008cc,Jaworski:2010dy,Avery:2012bj,Meier:2013fa} and on NM/FMI \cite{Uchida:2010jb} as well as pure \(\nabla T_\text{z}\) on NM/FMM \cite{Meier:2013dz} and on NM/FMI \cite{Uchida:2010jba,Uchida:2010fba,Huang:2012tk,Weiler:2012ki,Qu:2013jj,Kehlberger:2014ha,Siegel:2014dx,Agrawal:2014ik}. As summarized in Fig.~\ref{fig:parasiticeffects}~(c) an unintended \(\nabla T_\text{z}\) can hamper the evaluation of TSSE experiments with applied \(\nabla T_\text{x}\). Heat flow into the surrounding area or through the electrical contacts can induce an additional ANE in NM/FMM bilayers and NM/magnetic semiconductors as discussed in literature \cite{Bosu:2011bw,Huang:2011cd,Schmid:2013m,Meier:2013fa,Bui:2014ab,Soldatov:2014ab}. But since, in principle, all the effects of an LSSE experiment can be present in the TSSE experiment with unintended \(\nabla T_\text{z}\), proximity Nernst effects and especially parasitic LSSE can also be present in NM/FMI bilayers as already mentioned recently \cite{Schreier:2013ab}. This leads to four possible effects which are antisymmetric with respect to the external magnetic field, when the temperature gradient is not controlled very carefully (cf. Fig.~\ref{fig:parasiticeffects}~(c)). These phenomena and discussions of side-effects in TSSE experiments have not been treated systematically in the literature up to now for NM/FMI bilayers and will be investigated in this study.

The YIG films in our experiments had a thickness of \(t_\text{YIG}=\SI{180}{\nm}\) and were deposited on gadolinium gallium garnet (\ce{Gd3Ga5O12}, GGG) (111)-oriented single crystal substrates with width and length \(w=l=\SI{5}{\milli\meter}\) by pulsed laser deposition (PLD) from a stoichiometric polycrystalline target. The films show a coercive field of about \SI{100}{Oe} and a saturation magnetization of \(M_\text{S}=\SI[per-mode=symbol]{120}{\kilo\ampere\per\meter}\). The \ce{NFO} films with a thickness of about \(t_\text{NFO}=\SI{1}{\micro\meter}\) were deposited on \SI[product-units = power]{10 x 5}{\mm} \ce{MgAl2O4} (\ce{MAO}) (100)-oriented substrates by direct liquid injection chemical vapor deposition (DLI-CVD)\cite{Li:2011ka,Meier:2013dz}. After a vacuum break and cleaning with ethanol in an ultrasonic bath a \(t_\text{Pt}=\SI{10}{\nm}\) thin Pt strip was deposited by dc magnetron sputtering in an Ar atmosphere of \SI{1.5e-3}{\milli \bar} through a \SI{100}{\micro\meter} wide split-mask on one sample side of the \ce{YIG} and \ce{NFO} films with a length of \(l_\text{Pt}=\SI{5}{\milli\meter}\).

\begin{figure}[!h]
\centering
\includegraphics[width=\linewidth]{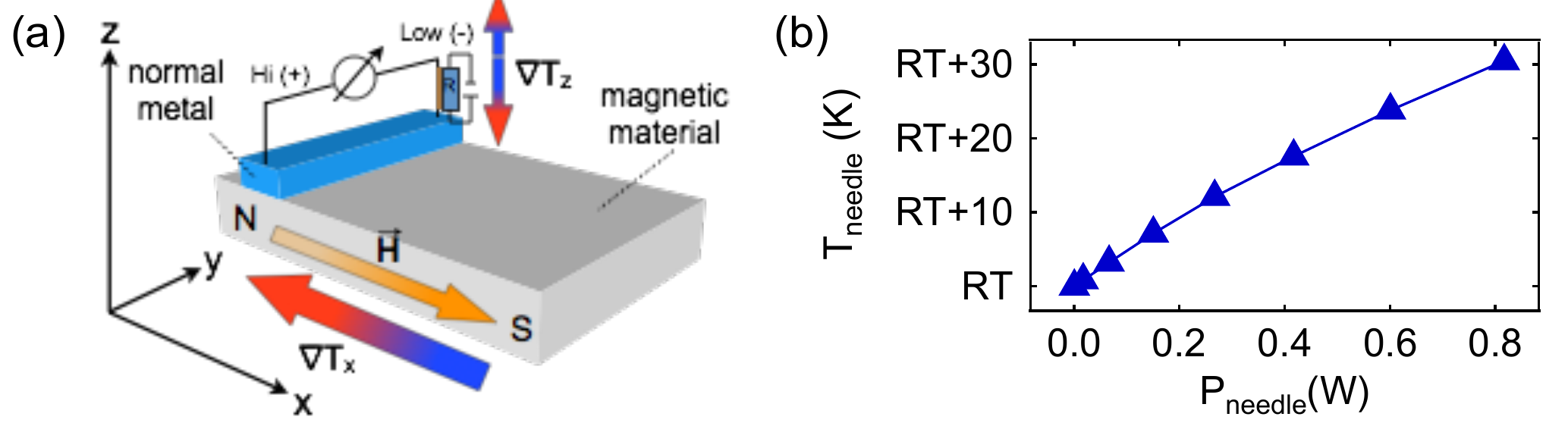}
\caption{(a) Measurement configuration: In-plane temperature gradient \(\nabla T_\text{x}\) applied parallel to the external magnetic field \(\vec{H}\) in the x-direction. An additional out-of-plane temperature gradient \(\nabla T_\text{z}\) can be induced by using thick contact tips or heating one tip with a voltage on a resistor \(R\) attached to the tip. (b) the temperature \(T_{\text{needle}}\) at the tip of the Au needle as a function of the power \(P_{\text{needle}}\) at the resistor \(R\) (only thermally connected to the contact needle, not electrically).}
\label{fig:setup_AuNeedle}
\end{figure}

For the experiments we used the same setup and technique described in Ref.\,\cite{Meier:2013fa}. The experiments are conducted under ambient conditions and the ends of the Pt strip were contacted with a microprobe system with Au and W tips of different diameters. Furthermore, one set of Au tips was equipped with a \SI{1.5}{\kilo\ohm} resistor for heating the tip to intentionally induce a \(\nabla T_\text{z}\) (cf. Fig.~\ref{fig:setup_AuNeedle}~(a)) \cite{Meier:2013fa}. Fig.~\ref{fig:setup_AuNeedle}~(b) shows the nearly linear relation between the power \(P_{\text{needle}}\) dissipated in the resistor and the tip temperature \(T_{\text{needle}}\) as determined with a type-K thermocouple glued to the tip in a calibration measurement. The voltage \(V\) at the Pt strip was measured with a Keithley 2182A nanovoltmeter. An external magnetic field \(H\) was applied along x in a range of \SI{+-600}{Oe} for \ce{YIG} and of \SI{+-1000}{Oe} for \ce{NFO} films.

\begin{figure}[!h]
\centering
\includegraphics[width=\linewidth]{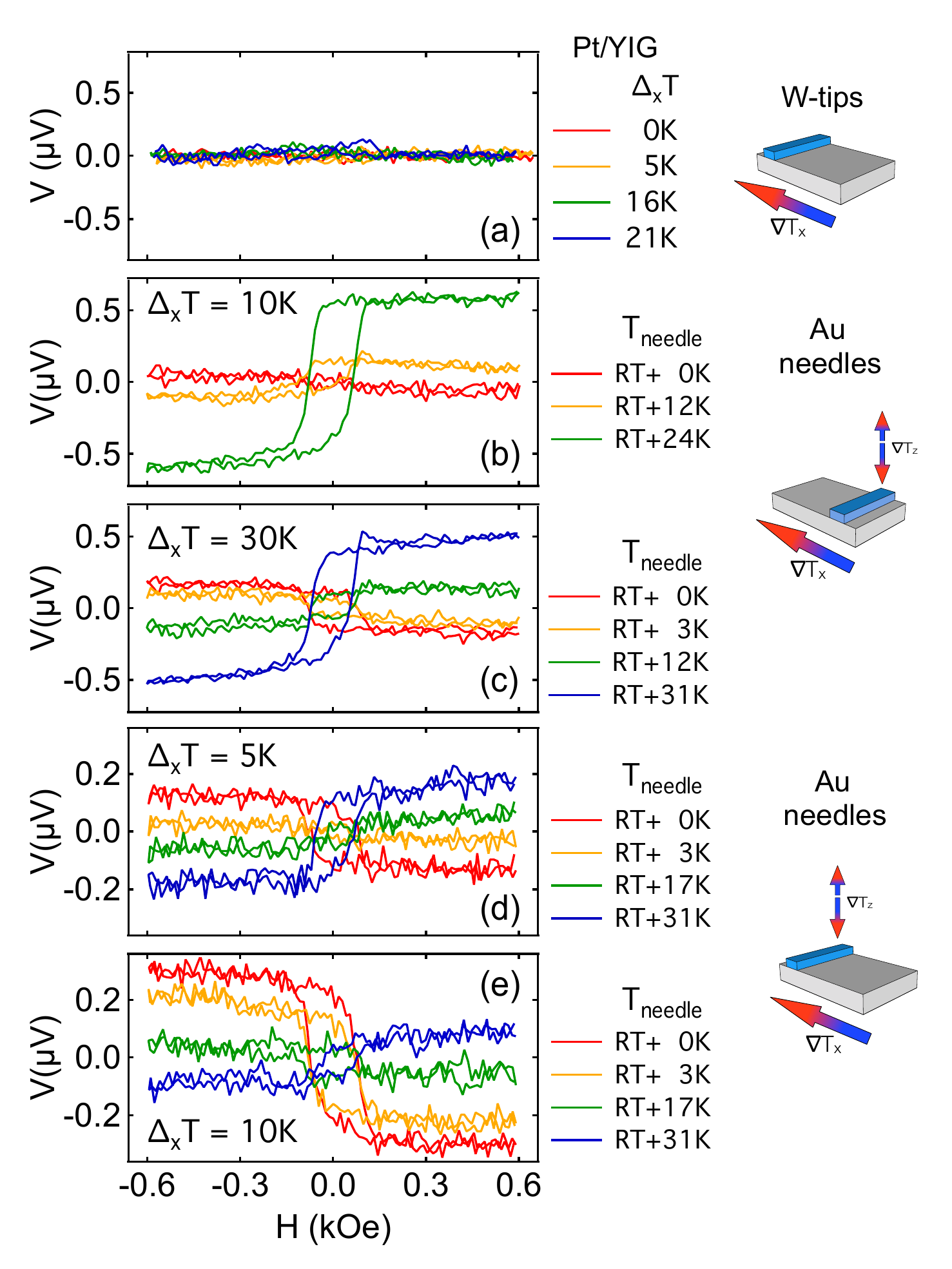}
\caption{(a) \(H\) dependence of \(V\) measured at the Pt strip on YIG located (a) at the hot side using thin W tips without inducing an additional \(\nabla T_\text{z}\) for various \(\nabla T_\text{x}\), (b)-(e) using thick Au tips and different tip heating voltages (b) at the cold side for \(\Delta T_\text{x} = \SI{10}{\kelvin}\), (c) at the cold side for \(\Delta T_\text{x} = \SI{30}{\kelvin}\), (d) at the hot side for \(\Delta T_\text{x} = \SI{5}{\kelvin}\) and (e) at the hot side for \(\Delta T_\text{x} = \SI{10}{\kelvin}\).}
\label{fig:YIG_W_Au_hot_cold}
\end{figure}

First, the ISHE voltage from the Pt/YIG sample was measured for various \(\Delta T_\text{x}\). The Pt strip was on the hot side and contacted with W tips. The voltage shows no significant variation within the sensitivity limit when \(H\) is varied (Fig.~\ref{fig:YIG_W_Au_hot_cold}~(a)). No Nernst effects are observed due to the insulating magnetic layer and no evidence for an additional \(\nabla T_\text{z}\) can be detected. Therefore, the clamping and heating of the sample in our setup and the contacting with thin W tips results in a pure \(\nabla T_\text{x}\) as already shown in Ref.\,\cite{Meier:2013fa}. TSSE is not observable although the Pt strip is located near the hot side of the YIG film and far away from the center where the TSSE should vanish.

In the next step we created a second heat sink by using thicker Au tips (Fig.~\ref{fig:YIG_W_Au_hot_cold}~(b)). The diameter of the contact area is of about \SI{500}{\micro\meter} in contrast to the contact area with W tips of about \SI{10}{\micro\meter}. It can be seen that the measured voltage now shows an antisymmetric behavior with respect to magnetic field inversion (Fig.~\ref{fig:YIG_W_Au_hot_cold}~(b)-(e)). Next, we applied a voltage to the Au tip resistor to heat the needle and change the out-of-plane heat flow. The heated Au needle is labeled \(T_{\text{needle}}\)\,=\,RT\,+\,x with room temperature RT\,=\,\SI{296}{\kelvin}. The magnitude \(V_{\text{sat}}\) (the voltage in saturation) is calculated with \((V_+ - V_-)/2\) for the average voltage \(V_\pm\) in the region of \(\SI{+500}{Oe} < H < \SI{+600}{Oe}\) and \(\SI{-600}{Oe} < H < \SI{-500}{Oe}\) for Pt/YIG as well as \(\SI{+900}{Oe} < H < \SI{+1000}{Oe}\) and \(\SI{-1000}{Oe} < H < \SI{-900}{Oe}\) for Pt/NFO, respectively.

In Fig.~\ref{fig:YIG_W_Au_hot_cold}~(b) for \(T_{\text{needle}} = \text{RT}\) a small antisymmetric effect of about \(V_{\text{sat}}=\SI{-50}{\nano\volt}\) is obtained when the Pt strip is at the cold side of the YIG film. When the needle is heated to \(T_{\text{needle}} = \text{RT}+\SI{12}{\kelvin}\) the ISHE voltage changes its sign to a value of \(V_\text{sat}=\SI[retain-explicit-plus]{+95}{\nano\volt}\) and changes further to \(V_\text{sat}=\SI[retain-explicit-plus]{+590}{\nano\volt}\) for \(T_{\text{needle}} = \text{RT}+\SI{24}{\kelvin}\). By using these Au needles with large contact areas an additional small out-of-plane heat flow is generated even at the cold side of the sample. This heat flow changes its sign with increasing \(T_{\text{needle}}\) which can be detected by the sign reversal of the measured voltage. When \(\nabla T_\text{x}\) is increased (\(\Delta T_\text{x} = \SI{30}{\kelvin}\) in Fig.~\ref{fig:YIG_W_Au_hot_cold}~(c)) the ISHE voltage at the Pt is \(V_\text{sat}=\SI{-170}{\nano\volt}\) for \(T_{\text{needle}}=\text{RT}\) and therefore three times larger than for \(\Delta T_\text{x} = \SI{10}{\kelvin}\). The ISHE voltage again increases with increasing \(T_{\text{needle}}\) and changes sign.

\begin{figure}[!h]
\centering
\includegraphics[width=\linewidth]{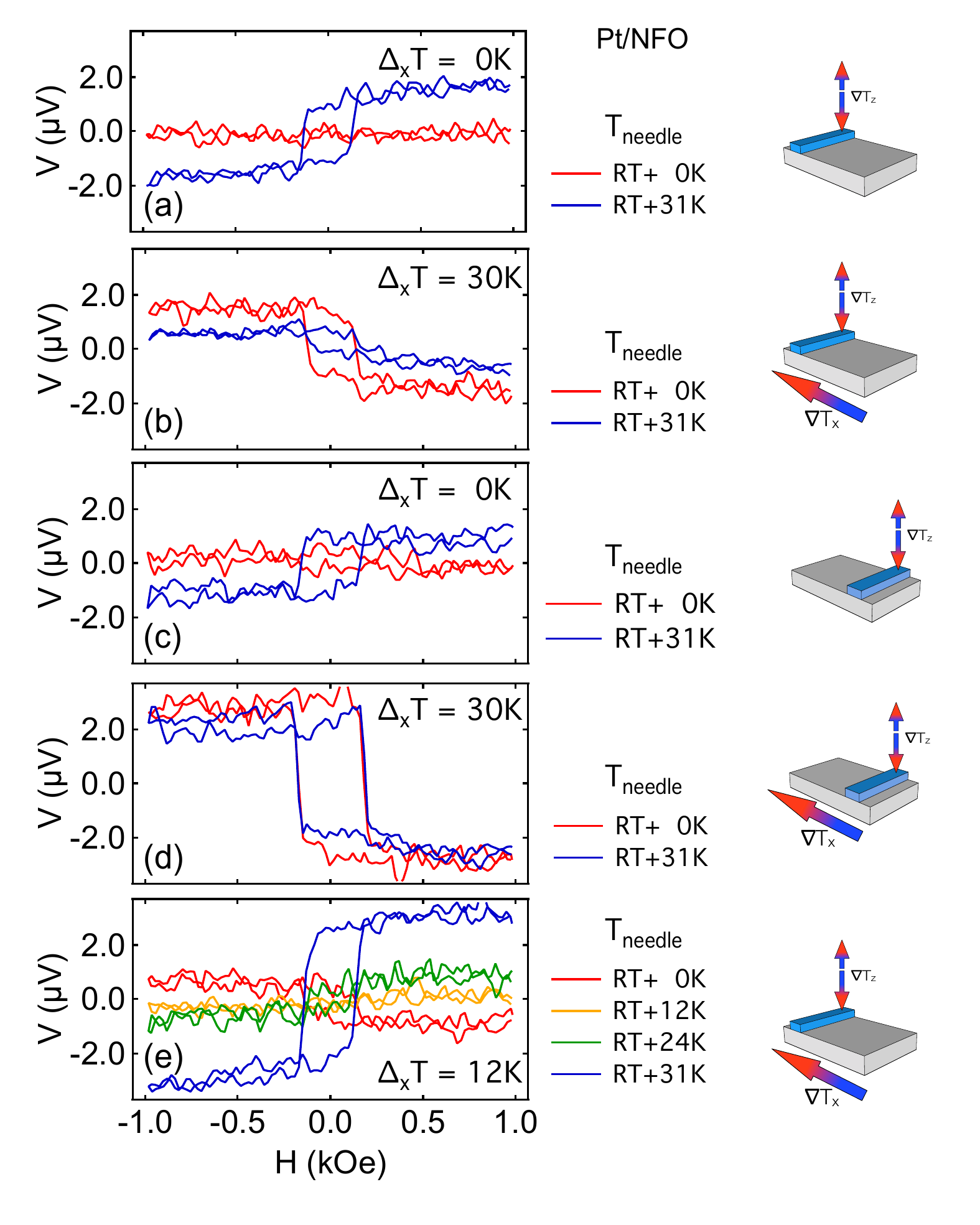}
\caption{(a)-(d) \(V\) as a function of \(H\) measured at a Pt strip on NFO using thick Au needles (a) at the hot side for \(\Delta T_\text{x} = \SI{0}{\kelvin}\), (b) at the hot side for \(\Delta T_\text{x} = \SI{30}{\kelvin}\), (c) at the cold side for \(\Delta T_\text{x} = \SI{0}{\kelvin}\), (d) at the cold side for \(\Delta T_\text{x} = \SI{30}{\kelvin}\), and (e) at the hot side with \(\Delta T_\text{x} = \SI{12}{\kelvin}\) and different Au tip heating voltages.}
\label{fig:NFO_Au_hot_cold}
\end{figure}

For a Pt strip at the hot side \(V_{\text{sat}}\) without tip heating is larger than at the cold side. For \(\Delta T_\text{x} = \SI{5}{\kelvin}\) the magnitude is about \(V_{\text{sat}}=\SI{-130}{\nano\volt}\) (Fig.~\ref{fig:YIG_W_Au_hot_cold}~(d)) which can be decreased to \(V_{\text{sat}}=\SI{-300}{\nano\volt}\) for \(\Delta T_\text{x} = \SI{10}{\kelvin}\) (Fig.~\ref{fig:YIG_W_Au_hot_cold}~(e)). The sign and the magnitude of \(V_{\text{sat}}\) can also be controlled by \(T_\text{needle}\) and, therefore, by \(\nabla T_\text{z}\). When \(T_\text{needle}\) is fixed at \(\text{RT}+\SI{31}{\kelvin}\), \(V_{\text{sat}}\) is about \SI[retain-explicit-plus]{+180}{\nano\volt} for \(\Delta T_\text{x} = \SI{5}{\kelvin}\) and \SI[retain-explicit-plus]{+90}{\nano\volt} for \(\Delta T_\text{x} = \SI{10}{\kelvin}\).

For a verification of this behavior in another material system, \ce{NFO} films with a Pt strip were used and contacted with Au needles. When \(\Delta T_\text{x} = \SI{0}{\kelvin}\) and no tip heating is applied no significant change in \(V\) as a function of \(H\) is observed (Fig.~\ref{fig:NFO_Au_hot_cold}~(a)). This behavior is the same for Pt strips on both sides of the \ce{NFO} film (Fig.~\ref{fig:NFO_Au_hot_cold}~(c)). 

When a \(\nabla T_\text{x}\) is applied, the same sign of \(V_{\text{sat}}\) on both sides is achieved (Fig.~\ref{fig:NFO_Au_hot_cold}~(b),(d)). The effect can again be manipulated by applying a needle temperature of \(T_{\text{needle}}=\text{RT}+\SI{31}{\kelvin}\). The discrepancy for Pt/NFO measurements on hot side (smaller \(|V_{\text{sat}}|\), Fig.~\ref{fig:NFO_Au_hot_cold}~(b)) and cold side (larger \(|V_{\text{sat}}|\), Fig.~\ref{fig:NFO_Au_hot_cold}~(d)) compared to Pt/YIG on hot side (larger \(|V_{\text{sat}}|\), Fig.~\ref{fig:YIG_W_Au_hot_cold}~(e)) and cold side (smaller \(|V_{\text{sat}}|\), Fig.~\ref{fig:YIG_W_Au_hot_cold}~(b)) can be explained by contacting the needles again after reversing the sample to "move" the Pt strip from the hot to the cold side. The real contact area between the tips and the Pt can be different when the sample is remounted. Furthermore, \(T_\text{needle}\) was varied for \(\Delta T_\text{x} = \SI{12}{\kelvin}\) (Fig.~\ref{fig:NFO_Au_hot_cold}~(e)). \(V_\text{sat}\) at \(T_\text{needle} = \text{RT}\) is about \SI[retain-explicit-plus]{+690}{\nano\volt}. The absolute value is more than two times smaller than for \(\Delta T_\text{x} = \SI{30}{\kelvin}\) (\(V_{\text{sat}}=\SI{-1550}{\nano\volt}\)) with the change of \(\Delta T_\text{x}\). \(V_\text{sat}\) also vanishes for \(T_\text{needle} = \text{RT}+\SI{12}{\kelvin}\) and changes sign for increasing \(T_\text{needle}\).

\begin{figure}[!h]
\centering
\includegraphics[width=\linewidth]{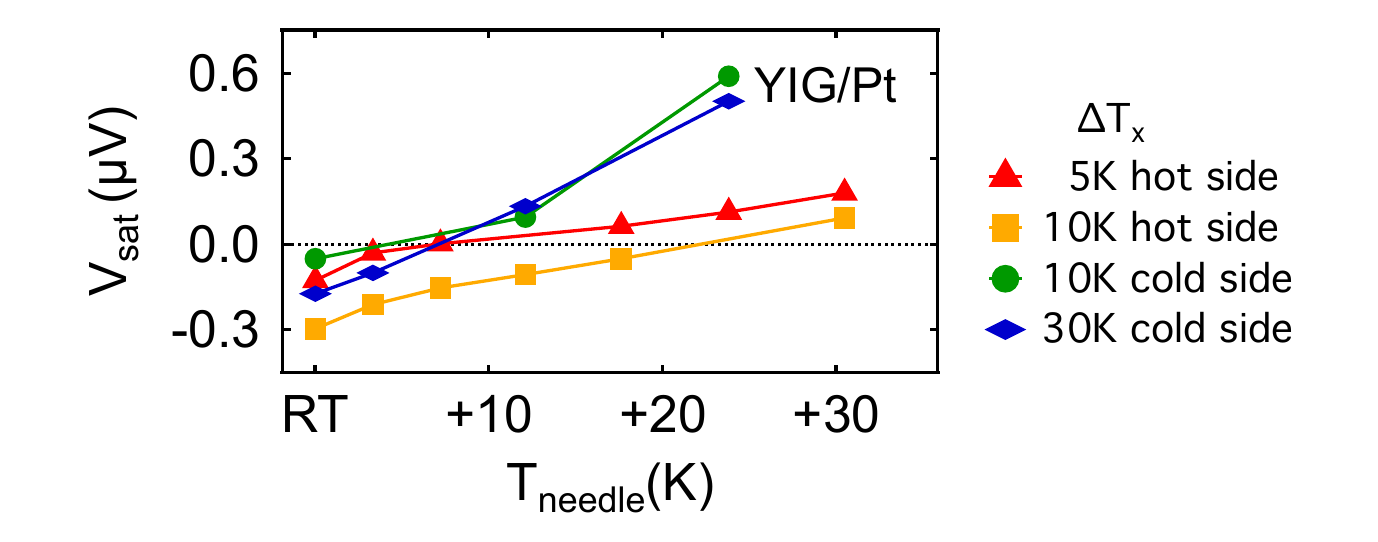}
\caption{\(V_{\text{sat}}\) as a function of the Au needle temperature \(T_{\text{needle}}\) for various in-plane temperature differences \(\Delta T_\text{x}\) in Pt/YIG.}
\label{fig:YIG_magnitude}
\end{figure}

\(V_{\text{sat}}\) measured for Pt/YIG at the hot and cold side is plotted as a function of \(T_\text{needle}\) for different \(\Delta T_\text{x}\) in Fig.~\ref{fig:YIG_magnitude}. A non-heated Au needle results in the same sign of \(V_{\text{sat}}\) for all \(\Delta T_\text{x}\), while \(|V_{\text{sat}}|\) is smaller on the cold side compared to the hot side. We explain this behaviour of \(V_\text{sat}\) by an unintended heat flux through the Au needles leading to a vertical temperature gradient \(\nabla T_\text{z}\) and, therefore, to a LSSE induced spin current into the Pt. We point out, that the sign of the resulting voltage at the Pt strip is quantitatively consistent with the magnitude of the recent LSSE reported by Schreier et al. \cite{Schreier:2014ax} from comparative experiments performed in different groups if we consider an unintended out-of-plane \(\nabla T_\text{z}\) pointing into -z direction. For a heated Au needle \(V_{\text{sat}}\) increases and crosses zero (cf. Fig.~\ref{fig:YIG_magnitude}). Here, the tip heating compensates the out-of-plane heat flux induced by \(\Delta T_\text{x}\) (\(\nabla T_\text{z}=0\)). After the sign change of \(V_{\text{sat}}\) (and therefore, \(\nabla T_\text{z}\)), the values increases with a larger (smaller) slope for the cold (hot) side. 

Xiao et al. \cite{Xiao:2010iy} discussed the temperature difference \(\Delta T_\text{me}\) between the magnon temperature in the FM and the electron temperature in the NM as the origin of the thermally induced spin current. \(\Delta T_\text{me}\) can be inferred from the recorded voltage as \cite{Xiao:2010iy,Schreier:2013ab}

\begin{equation}
	\Delta T_\text{me} = \frac{V_\text{sat}\,\pi\,M_\text{S}\,V_\text{a}\,t_\text{Pt}}{g_\text{r}\,\gamma\,k_\text{B}\,e\,\Theta_\text{SH}\,\rho_\text{Pt}\,l_\text{Pt}\,\lambda\,\tanh(t/2\lambda)}.
\label{magnonelectrontemperature}
\end{equation}

Here, \(V_\text{a}\) is the magnetic coherence volume, \(g_\text{r}\) is the real part of the spin mixing conductance, \(\gamma\) is the gyromagnetic ratio, \(k_\text{B}\) is the Boltzmann constant, \(e\) is the elementary charge, \(\Theta_\text{SH}\) is the spin Hall angle, and \(\lambda\) is the spin diffusion length of the NM material. The two temperature model has proven successful in relating \(\Delta T_\text{me}\) to the phonon temperature, accessible in experiments \cite{Schreier:2013ab,Flipse:2012kn}. We simulate the phonon and magnon temperatures assuming 1D transport in our films and disregard the influence of thermal contact resistance other than the coupling between magnons and electrons. This yields a value \(\Delta T_\text{z}\), the (phonon) temperature drop across the YIG film, for which the experimentally measured \(V_\text{sat}\) is obtained. We assume \(\rho_\text{Pt} = \SI{40}{\micro\ohm\centi\meter}\). All material dependent YIG parameters were taken from Ref.\,\cite{Schreier:2013ab}.

We calculated \(\Delta T_\text{me}\) for the largest and smallest \(V_\text{sat}\) taken from Fig.~\ref{fig:YIG_magnitude} at RT (red curves in Fig.~\ref{fig:YIG_W_Au_hot_cold}~(b) and (e)) and the corresponding \(\Delta T_\text{z}\). For \(|V_\text{sat}|=\SI{50}{\nano\volt}\) (cf. Fig.~\ref{fig:YIG_W_Au_hot_cold}~(b)) we obtain \(\Delta T_\text{me}=\SI{0.5}{\micro\kelvin}\) and a corresponding \(\Delta T_\text{z}=\SI{2}{\milli\kelvin}\). For \(|V_\text{sat}|=\SI{300}{\nano\volt}\) (cf. Fig.~\ref{fig:YIG_W_Au_hot_cold}~(e)) the corresponding values are \(\Delta T_\text{me}=\SI{3.2}{\micro\kelvin}\) and \(\Delta T_\text{z}=\SI{12}{\milli\kelvin}\). The obtained \(\Delta T_\text{z}\) are in the order of a few millikelvins. It is reasonable to assume that such values can be induced by e.g. thick contact tips, especially considering that our initial simplifications should lead to an overestimation of \(\Delta T_\text{z}\) \cite{Schreier:2013ab}. The transverse spin Seebeck configuration was investigated in detail in Ref.\,\cite{Schreier:2013ab}. It was found that for \(\Delta T_\text{x} = \SI{20}{\kelvin}\) the obtained \(\Delta T_\text{me}\) is well below \SI{1}{\micro\kelvin}, even at the very edge of the sample where \(\Delta T_\text{me}\) is maximized. This further supports the notion that spurious out-of-plane gradients are responsible for the voltages observed in our samples.

The different magnitudes in voltage for Pt/NFO compared to Pt/YIG can be explained by different contact areas of the tips, different thicknesses and thermal conductivities between NFO and YIG as well as different spin mixing conductances \cite{Weiler:2013ko,Qiu:2013je}. However, for both sample systems the spin mixing conductance should be large enough to observe a thermally driven spin current across the NM/FMI interface, since we clearly observe an LSSE due to the Au tip heating induced \(\Delta T_\text{z}\).

In addition to the LSSE, we now discuss other parasitic effects like the ANE and proximity ANE, which can be produced by an unintended \(\Delta T_\text{z}\) (see Fig.~\ref{fig:parasiticeffects}). We can exclude an ANE for YIG due to the lack of charge carriers. For NFO we observed an ANE which is one order of magnitude smaller at RT than the LSSE \cite{Meier:2013dz}. This ANE can be explained by the weak conductance of NFO at RT due to thermal activation energies of a few hundred meV depending on the preparation technique \cite{Meier:2013dz,Klewe:2014km}. The proximity ANE in Pt/NFO can also be excluded, since no spin polarization was found using x-ray resonant magnetic reflectivity (XRMR) measurements, which are very sensitive to the interface spin polarization \cite{Kuschel:2014ab}. In case of Pt/YIG Gepr\"ags et al. presented x-ray magnetic circular dichroism measurements (XMCD) with no evidence for any spin polarization in Pt \cite{Geprags:2012be}, while Lu et al. could show XMCD measurements indicating magnetic moments in Pt on their YIG samples \cite{Lu:2013eq}. Future investigations with XRMR can give more insight to this discrepancy. However, Kikkawa et al. could show that a potential contribution of a proximity ANE additional to the LSSE is negligibly small \cite{Kikkawa:2013hx}. This supports our conclusion, that the main antisymmetric contribution in our measurements on both Pt/YIG and Pt/NFO is the LSSE, which is driven by an out-of-plane temperature gradient.

We do not observe any symmetric contribution for \(\nabla T_\text{x}\) without tip heating. Therefore, PNE and proximity PNE contributions can also be excluded. Nevertheless, we find a small symmetric contribution for strong tip heating as demonstrated in Fig.~\ref{fig:YIG_W_Au_hot_cold}~(c) for \(\Delta T_{\text{needle}} = \text{RT} + \SI{31}{\kelvin}\). In the region of \(H_C\) small peaks are visible under symmetrization of the voltage. This hints at the existence of an additional magnetothermopower effect potentially induced by a temperature gradient \(\nabla T_\text{y}\) and will be part of future investigations.

Recently, Wegrowe et al. \cite{Wegrowe:2014bs} used anisotropic heat-transport as an interpretation for the measured voltages using in-plane temperature gradients. In their work, they derived the anisotropic field-dependent temperature gradient in FMM and FMI from the Onsager reciprocity relations. Therefore, the thermocouple effect between the FM, the NM and the contacting tips can generate field-dependent voltages if there is a difference in the Seebeck coefficients. In our investigated systems, the Seebeck coefficients are indeed different for FM, Pt and the contact tips. However, since we do not observe a field-dependent variation of the ISHE voltage by using W tips, any anisotropic field-dependent heat-transport can be excluded as the reason for the observed voltages.

In summary, we investigated the relevance of TSSE in Pt/YIG and Pt/NFO systems. We found no significant ISHE voltages upon applying an in-plane temperature gradient and using sharp W tips  (\SI{10}{\micro\meter} contact diameter) for the electrical contacting. However, upon using tips with more than \SI{100}{\micro\meter} diameter contact area, which induce an additional out-of-plane temperature gradient, an antisymmetric contribution to the ISHE voltage of the Pt strip could be introduced at will. This antisymmetric effect can be identified as LSSE which was verified by controlling the needle temperature and varying the out-of-plane temperature gradient. Taken together, in all our experiments, we thus only observe LSSE-type signatures. These LSSE voltages can be reminiscent of a TSSE-type response if an unintentional (or intentional) \(\nabla T_\text{z}\) is present. This shows that utmost care is required if one is to interpret magnetothermopower effects in terms of the TSSE.

We thank the DFG (SPP 1538) and the EMRP JRP EXL04 SpinCal for financial support. The EMRP is jointly funded by the EMRP participating countries within EURAMET and the EU.

\bibliographystyle{apsrev4-1}
\bibliography{bibfile}

\begin{thebibliography}{37}%
\makeatletter
\providecommand \@ifxundefined [1]{%
 \@ifx{#1\undefined}
}%
\providecommand \@ifnum [1]{%
 \ifnum #1\expandafter \@firstoftwo
 \else \expandafter \@secondoftwo
 \fi
}%
\providecommand \@ifx [1]{%
 \ifx #1\expandafter \@firstoftwo
 \else \expandafter \@secondoftwo
 \fi
}%
\providecommand \natexlab [1]{#1}%
\providecommand \enquote  [1]{``#1''}%
\providecommand \bibnamefont  [1]{#1}%
\providecommand \bibfnamefont [1]{#1}%
\providecommand \citenamefont [1]{#1}%
\providecommand \href@noop [0]{\@secondoftwo}%
\providecommand \href [0]{\begingroup \@sanitize@url \@href}%
\providecommand \@href[1]{\@@startlink{#1}\@@href}%
\providecommand \@@href[1]{\endgroup#1\@@endlink}%
\providecommand \@sanitize@url [0]{\catcode `\\12\catcode `\$12\catcode
  `\&12\catcode `\#12\catcode `\^12\catcode `\_12\catcode `\%12\relax}%
\providecommand \@@startlink[1]{}%
\providecommand \@@endlink[0]{}%
\providecommand \url  [0]{\begingroup\@sanitize@url \@url }%
\providecommand \@url [1]{\endgroup\@href {#1}{\urlprefix }}%
\providecommand \urlprefix  [0]{URL }%
\providecommand \Eprint [0]{\href }%
\providecommand \doibase [0]{http://dx.doi.org/}%
\providecommand \selectlanguage [0]{\@gobble}%
\providecommand \bibinfo  [0]{\@secondoftwo}%
\providecommand \bibfield  [0]{\@secondoftwo}%
\providecommand \translation [1]{[#1]}%
\providecommand \BibitemOpen [0]{}%
\providecommand \bibitemStop [0]{}%
\providecommand \bibitemNoStop [0]{.\EOS\space}%
\providecommand \EOS [0]{\spacefactor3000\relax}%
\providecommand \BibitemShut  [1]{\csname bibitem#1\endcsname}%
\let\auto@bib@innerbib\@empty
\bibitem [{\citenamefont {Bauer}\ \emph {et~al.}(2012)\citenamefont {Bauer},
  \citenamefont {Saitoh},\ and\ \citenamefont {van Wees}}]{Bauer:2012fq}%
  \BibitemOpen
  \bibfield  {author} {\bibinfo {author} {\bibfnamefont {G.~E.~W.}\
  \bibnamefont {Bauer}}, \bibinfo {author} {\bibfnamefont {E.}~\bibnamefont
  {Saitoh}}, \ and\ \bibinfo {author} {\bibfnamefont {B.~J.}\ \bibnamefont {van
  Wees}},\ }\href@noop {} {\bibfield  {journal} {\bibinfo  {journal} {Nature
  Materials}\ }\textbf {\bibinfo {volume} {11}},\ \bibinfo {pages} {391}
  (\bibinfo {year} {2012})}\BibitemShut {NoStop}%
\bibitem [{\citenamefont {Wolf}\ \emph {et~al.}(2001)\citenamefont {Wolf},
  \citenamefont {Awschalom}, \citenamefont {Buhrman}, \citenamefont {Daughton},
  \citenamefont {von Moln\'{a}r}, \citenamefont {Roukes}, \citenamefont
  {Chtchelkanova},\ and\ \citenamefont {Treger}}]{Wolf:2001fu}%
  \BibitemOpen
  \bibfield  {author} {\bibinfo {author} {\bibfnamefont {S.~A.}\ \bibnamefont
  {Wolf}}, \bibinfo {author} {\bibfnamefont {D.~D.}\ \bibnamefont {Awschalom}},
  \bibinfo {author} {\bibfnamefont {R.~A.}\ \bibnamefont {Buhrman}}, \bibinfo
  {author} {\bibfnamefont {J.~M.}\ \bibnamefont {Daughton}}, \bibinfo {author}
  {\bibfnamefont {S.}~\bibnamefont {von Moln\'{a}r}}, \bibinfo {author}
  {\bibfnamefont {M.~L.}\ \bibnamefont {Roukes}}, \bibinfo {author}
  {\bibfnamefont {A.~Y.}\ \bibnamefont {Chtchelkanova}}, \ and\ \bibinfo
  {author} {\bibfnamefont {D.~M.}\ \bibnamefont {Treger}},\ }\href@noop {}
  {\bibfield  {journal} {\bibinfo  {journal} {Science}\ }\textbf {\bibinfo
  {volume} {294}},\ \bibinfo {pages} {1488} (\bibinfo {year}
  {2001})}\BibitemShut {NoStop}%
\bibitem [{\citenamefont {Kirihara}\ \emph {et~al.}(2012)\citenamefont
  {Kirihara}, \citenamefont {Uchida}, \citenamefont {Kajiwara}, \citenamefont
  {Ishida}, \citenamefont {Nakamura}, \citenamefont {Manako}, \citenamefont
  {Saitoh},\ and\ \citenamefont {Yorozu}}]{Kirihara:2012jq}%
  \BibitemOpen
  \bibfield  {author} {\bibinfo {author} {\bibfnamefont {A.}~\bibnamefont
  {Kirihara}}, \bibinfo {author} {\bibfnamefont {K.}~\bibnamefont {Uchida}},
  \bibinfo {author} {\bibfnamefont {Y.}~\bibnamefont {Kajiwara}}, \bibinfo
  {author} {\bibfnamefont {M.}~\bibnamefont {Ishida}}, \bibinfo {author}
  {\bibfnamefont {Y.}~\bibnamefont {Nakamura}}, \bibinfo {author}
  {\bibfnamefont {T.}~\bibnamefont {Manako}}, \bibinfo {author} {\bibfnamefont
  {E.}~\bibnamefont {Saitoh}}, \ and\ \bibinfo {author} {\bibfnamefont
  {S.}~\bibnamefont {Yorozu}},\ }\href@noop {} {\bibfield  {journal} {\bibinfo
  {journal} {Nature Materials}\ }\textbf {\bibinfo {volume} {11}},\ \bibinfo
  {pages} {686} (\bibinfo {year} {2012})}\BibitemShut {NoStop}%
\bibitem [{\citenamefont {Uchida}\ \emph {et~al.}(2008)\citenamefont {Uchida},
  \citenamefont {Takahashi}, \citenamefont {Harii}, \citenamefont {Ieda},
  \citenamefont {Koshibae}, \citenamefont {Ando}, \citenamefont {Maekawa},\
  and\ \citenamefont {Saitoh}}]{Uchida:2008cc}%
  \BibitemOpen
  \bibfield  {author} {\bibinfo {author} {\bibfnamefont {K.}~\bibnamefont
  {Uchida}}, \bibinfo {author} {\bibfnamefont {S.}~\bibnamefont {Takahashi}},
  \bibinfo {author} {\bibfnamefont {K.}~\bibnamefont {Harii}}, \bibinfo
  {author} {\bibfnamefont {J.}~\bibnamefont {Ieda}}, \bibinfo {author}
  {\bibfnamefont {W.}~\bibnamefont {Koshibae}}, \bibinfo {author}
  {\bibfnamefont {K.}~\bibnamefont {Ando}}, \bibinfo {author} {\bibfnamefont
  {S.}~\bibnamefont {Maekawa}}, \ and\ \bibinfo {author} {\bibfnamefont
  {E.}~\bibnamefont {Saitoh}},\ }\href@noop {} {\bibfield  {journal} {\bibinfo
  {journal} {Nature}\ }\textbf {\bibinfo {volume} {455}},\ \bibinfo {pages}
  {778} (\bibinfo {year} {2008})}\BibitemShut {NoStop}%
\bibitem [{\citenamefont {Saitoh}\ \emph {et~al.}(2006)\citenamefont {Saitoh},
  \citenamefont {Ueda}, \citenamefont {Miyajima},\ and\ \citenamefont
  {Tatara}}]{Saitoh:2006kk}%
  \BibitemOpen
  \bibfield  {author} {\bibinfo {author} {\bibfnamefont {E.}~\bibnamefont
  {Saitoh}}, \bibinfo {author} {\bibfnamefont {M.}~\bibnamefont {Ueda}},
  \bibinfo {author} {\bibfnamefont {H.}~\bibnamefont {Miyajima}}, \ and\
  \bibinfo {author} {\bibfnamefont {G.}~\bibnamefont {Tatara}},\ }\href@noop {}
  {\bibfield  {journal} {\bibinfo  {journal} {Applied Physics Letters}\
  }\textbf {\bibinfo {volume} {88}},\ \bibinfo {pages} {182509} (\bibinfo
  {year} {2006})}\BibitemShut {NoStop}%
\bibitem [{\citenamefont {Avery}\ \emph {et~al.}(2012)\citenamefont {Avery},
  \citenamefont {Pufall},\ and\ \citenamefont {Zink}}]{Avery:2012bj}%
  \BibitemOpen
  \bibfield  {author} {\bibinfo {author} {\bibfnamefont {A.}~\bibnamefont
  {Avery}}, \bibinfo {author} {\bibfnamefont {M.}~\bibnamefont {Pufall}}, \
  and\ \bibinfo {author} {\bibfnamefont {B.}~\bibnamefont {Zink}},\ }\href@noop
  {} {\bibfield  {journal} {\bibinfo  {journal} {Physical Review Letters}\
  }\textbf {\bibinfo {volume} {109}},\ \bibinfo {pages} {196602} (\bibinfo
  {year} {2012})}\BibitemShut {NoStop}%
\bibitem [{\citenamefont {Uchida}\ \emph
  {et~al.}(2010{\natexlab{a}})\citenamefont {Uchida}, \citenamefont {Adachi},
  \citenamefont {Ota}, \citenamefont {Nakayama}, \citenamefont {Maekawa},\ and\
  \citenamefont {Saitoh}}]{Uchida:2010jb}%
  \BibitemOpen
  \bibfield  {author} {\bibinfo {author} {\bibfnamefont {K.}~\bibnamefont
  {Uchida}}, \bibinfo {author} {\bibfnamefont {H.}~\bibnamefont {Adachi}},
  \bibinfo {author} {\bibfnamefont {T.}~\bibnamefont {Ota}}, \bibinfo {author}
  {\bibfnamefont {H.}~\bibnamefont {Nakayama}}, \bibinfo {author}
  {\bibfnamefont {S.}~\bibnamefont {Maekawa}}, \ and\ \bibinfo {author}
  {\bibfnamefont {E.}~\bibnamefont {Saitoh}},\ }\href@noop {} {\bibfield
  {journal} {\bibinfo  {journal} {Applied Physics Letters}\ }\textbf {\bibinfo
  {volume} {97}},\ \bibinfo {pages} {172505} (\bibinfo {year}
  {2010}{\natexlab{a}})}\BibitemShut {NoStop}%
\bibitem [{\citenamefont {Meier}\ \emph
  {et~al.}(2013{\natexlab{a}})\citenamefont {Meier}, \citenamefont {Kuschel},
  \citenamefont {Shen}, \citenamefont {Gupta}, \citenamefont {Kikkawa},
  \citenamefont {Uchida}, \citenamefont {Saitoh}, \citenamefont {Schmalhorst},\
  and\ \citenamefont {Reiss}}]{Meier:2013dz}%
  \BibitemOpen
  \bibfield  {author} {\bibinfo {author} {\bibfnamefont {D.}~\bibnamefont
  {Meier}}, \bibinfo {author} {\bibfnamefont {T.}~\bibnamefont {Kuschel}},
  \bibinfo {author} {\bibfnamefont {L.}~\bibnamefont {Shen}}, \bibinfo {author}
  {\bibfnamefont {A.}~\bibnamefont {Gupta}}, \bibinfo {author} {\bibfnamefont
  {T.}~\bibnamefont {Kikkawa}}, \bibinfo {author} {\bibfnamefont
  {K.}~\bibnamefont {Uchida}}, \bibinfo {author} {\bibfnamefont
  {E.}~\bibnamefont {Saitoh}}, \bibinfo {author} {\bibfnamefont {J.~M.}\
  \bibnamefont {Schmalhorst}}, \ and\ \bibinfo {author} {\bibfnamefont
  {G.}~\bibnamefont {Reiss}},\ }\href@noop {} {\bibfield  {journal} {\bibinfo
  {journal} {Physical Review B}\ }\textbf {\bibinfo {volume} {87}},\ \bibinfo
  {pages} {054421} (\bibinfo {year} {2013}{\natexlab{a}})}\BibitemShut
  {NoStop}%
\bibitem [{\citenamefont {Guo}\ \emph {et~al.}(2014)\citenamefont {Guo},
  \citenamefont {Niu},\ and\ \citenamefont {Nagaosa}}]{Guo:2014bz}%
  \BibitemOpen
  \bibfield  {author} {\bibinfo {author} {\bibfnamefont {G.~Y.}\ \bibnamefont
  {Guo}}, \bibinfo {author} {\bibfnamefont {Q.}~\bibnamefont {Niu}}, \ and\
  \bibinfo {author} {\bibfnamefont {N.}~\bibnamefont {Nagaosa}},\ }\href@noop
  {} {\bibfield  {journal} {\bibinfo  {journal} {Physical Review B}\ }\textbf
  {\bibinfo {volume} {89}},\ \bibinfo {pages} {214406} (\bibinfo {year}
  {2014})}\BibitemShut {NoStop}%
\bibitem [{\citenamefont {Jaworski}\ \emph {et~al.}(2010)\citenamefont
  {Jaworski}, \citenamefont {Yang}, \citenamefont {Mack}, \citenamefont
  {Awschalom}, \citenamefont {Heremans},\ and\ \citenamefont
  {Myers}}]{Jaworski:2010dy}%
  \BibitemOpen
  \bibfield  {author} {\bibinfo {author} {\bibfnamefont {C.~M.}\ \bibnamefont
  {Jaworski}}, \bibinfo {author} {\bibfnamefont {J.}~\bibnamefont {Yang}},
  \bibinfo {author} {\bibfnamefont {S.}~\bibnamefont {Mack}}, \bibinfo {author}
  {\bibfnamefont {D.~D.}\ \bibnamefont {Awschalom}}, \bibinfo {author}
  {\bibfnamefont {J.~P.}\ \bibnamefont {Heremans}}, \ and\ \bibinfo {author}
  {\bibfnamefont {R.~C.}\ \bibnamefont {Myers}},\ }\href@noop {} {\bibfield
  {journal} {\bibinfo  {journal} {Nature Materials}\ }\textbf {\bibinfo
  {volume} {9}},\ \bibinfo {pages} {898} (\bibinfo {year} {2010})}\BibitemShut
  {NoStop}%
\bibitem [{\citenamefont {Meier}\ \emph
  {et~al.}(2013{\natexlab{b}})\citenamefont {Meier}, \citenamefont {Reinhardt},
  \citenamefont {Schmid}, \citenamefont {Back}, \citenamefont {Schmalhorst},
  \citenamefont {Kuschel},\ and\ \citenamefont {Reiss}}]{Meier:2013fa}%
  \BibitemOpen
  \bibfield  {author} {\bibinfo {author} {\bibfnamefont {D.}~\bibnamefont
  {Meier}}, \bibinfo {author} {\bibfnamefont {D.}~\bibnamefont {Reinhardt}},
  \bibinfo {author} {\bibfnamefont {M.}~\bibnamefont {Schmid}}, \bibinfo
  {author} {\bibfnamefont {C.~H.}\ \bibnamefont {Back}}, \bibinfo {author}
  {\bibfnamefont {J.~M.}\ \bibnamefont {Schmalhorst}}, \bibinfo {author}
  {\bibfnamefont {T.}~\bibnamefont {Kuschel}}, \ and\ \bibinfo {author}
  {\bibfnamefont {G.}~\bibnamefont {Reiss}},\ }\href@noop {} {\bibfield
  {journal} {\bibinfo  {journal} {Physical Review B}\ }\textbf {\bibinfo
  {volume} {88}},\ \bibinfo {pages} {184425} (\bibinfo {year}
  {2013}{\natexlab{b}})}\BibitemShut {NoStop}%
\bibitem [{\citenamefont {Uchida}\ \emph
  {et~al.}(2010{\natexlab{b}})\citenamefont {Uchida}, \citenamefont {Adachi},
  \citenamefont {Ota}, \citenamefont {Nakayama}, \citenamefont {Maekawa},\ and\
  \citenamefont {Saitoh}}]{Uchida:2010jba}%
  \BibitemOpen
  \bibfield  {author} {\bibinfo {author} {\bibfnamefont {K.}~\bibnamefont
  {Uchida}}, \bibinfo {author} {\bibfnamefont {H.}~\bibnamefont {Adachi}},
  \bibinfo {author} {\bibfnamefont {T.}~\bibnamefont {Ota}}, \bibinfo {author}
  {\bibfnamefont {H.}~\bibnamefont {Nakayama}}, \bibinfo {author}
  {\bibfnamefont {S.}~\bibnamefont {Maekawa}}, \ and\ \bibinfo {author}
  {\bibfnamefont {E.}~\bibnamefont {Saitoh}},\ }\href@noop {} {\bibfield
  {journal} {\bibinfo  {journal} {Applied Physics Letters}\ }\textbf {\bibinfo
  {volume} {97}},\ \bibinfo {pages} {172505} (\bibinfo {year}
  {2010}{\natexlab{b}})}\BibitemShut {NoStop}%
\bibitem [{\citenamefont {Uchida}\ \emph
  {et~al.}(2010{\natexlab{c}})\citenamefont {Uchida}, \citenamefont {Nonaka},
  \citenamefont {Ota},\ and\ \citenamefont {Saitoh}}]{Uchida:2010fba}%
  \BibitemOpen
  \bibfield  {author} {\bibinfo {author} {\bibfnamefont {K.}~\bibnamefont
  {Uchida}}, \bibinfo {author} {\bibfnamefont {T.}~\bibnamefont {Nonaka}},
  \bibinfo {author} {\bibfnamefont {T.}~\bibnamefont {Ota}}, \ and\ \bibinfo
  {author} {\bibfnamefont {E.}~\bibnamefont {Saitoh}},\ }\href@noop {}
  {\bibfield  {journal} {\bibinfo  {journal} {Applied Physics Letters}\
  }\textbf {\bibinfo {volume} {97}},\ \bibinfo {pages} {262504} (\bibinfo
  {year} {2010}{\natexlab{c}})}\BibitemShut {NoStop}%
\bibitem [{\citenamefont {Huang}\ \emph {et~al.}(2012)\citenamefont {Huang},
  \citenamefont {Fan}, \citenamefont {Qu}, \citenamefont {Chen}, \citenamefont
  {Wang}, \citenamefont {Wu}, \citenamefont {Chen}, \citenamefont {Xiao},\ and\
  \citenamefont {Chien}}]{Huang:2012tk}%
  \BibitemOpen
  \bibfield  {author} {\bibinfo {author} {\bibfnamefont {S.~Y.}\ \bibnamefont
  {Huang}}, \bibinfo {author} {\bibfnamefont {X.}~\bibnamefont {Fan}}, \bibinfo
  {author} {\bibfnamefont {D.}~\bibnamefont {Qu}}, \bibinfo {author}
  {\bibfnamefont {Y.~P.}\ \bibnamefont {Chen}}, \bibinfo {author}
  {\bibfnamefont {W.~G.}\ \bibnamefont {Wang}}, \bibinfo {author}
  {\bibfnamefont {J.}~\bibnamefont {Wu}}, \bibinfo {author} {\bibfnamefont
  {T.~Y.}\ \bibnamefont {Chen}}, \bibinfo {author} {\bibfnamefont {J.~Q.}\
  \bibnamefont {Xiao}}, \ and\ \bibinfo {author} {\bibfnamefont {C.~L.}\
  \bibnamefont {Chien}},\ }\href@noop {} {\bibfield  {journal} {\bibinfo
  {journal} {Physical Review Letters}\ }\textbf {\bibinfo {volume} {109}},\
  \bibinfo {pages} {107204} (\bibinfo {year} {2012})}\BibitemShut {NoStop}%
\bibitem [{\citenamefont {Weiler}\ \emph {et~al.}(2012)\citenamefont {Weiler},
  \citenamefont {Althammer}, \citenamefont {Czeschka}, \citenamefont {Huebl},
  \citenamefont {Wagner}, \citenamefont {Opel}, \citenamefont {Imort},
  \citenamefont {Reiss}, \citenamefont {Thomas}, \citenamefont {Gross},\ and\
  \citenamefont {Goennenwein}}]{Weiler:2012ki}%
  \BibitemOpen
  \bibfield  {author} {\bibinfo {author} {\bibfnamefont {M.}~\bibnamefont
  {Weiler}}, \bibinfo {author} {\bibfnamefont {M.}~\bibnamefont {Althammer}},
  \bibinfo {author} {\bibfnamefont {F.}~\bibnamefont {Czeschka}}, \bibinfo
  {author} {\bibfnamefont {H.}~\bibnamefont {Huebl}}, \bibinfo {author}
  {\bibfnamefont {M.}~\bibnamefont {Wagner}}, \bibinfo {author} {\bibfnamefont
  {M.}~\bibnamefont {Opel}}, \bibinfo {author} {\bibfnamefont {I.-M.}\
  \bibnamefont {Imort}}, \bibinfo {author} {\bibfnamefont {G.}~\bibnamefont
  {Reiss}}, \bibinfo {author} {\bibfnamefont {A.}~\bibnamefont {Thomas}},
  \bibinfo {author} {\bibfnamefont {R.}~\bibnamefont {Gross}}, \ and\ \bibinfo
  {author} {\bibfnamefont {S.}~\bibnamefont {Goennenwein}},\ }\href@noop {}
  {\bibfield  {journal} {\bibinfo  {journal} {Physical Review Letters}\
  }\textbf {\bibinfo {volume} {108}},\ \bibinfo {pages} {106602} (\bibinfo
  {year} {2012})}\BibitemShut {NoStop}%
\bibitem [{\citenamefont {Qu}\ \emph {et~al.}(2013)\citenamefont {Qu},
  \citenamefont {Huang}, \citenamefont {Hu}, \citenamefont {Wu},\ and\
  \citenamefont {Chien}}]{Qu:2013jj}%
  \BibitemOpen
  \bibfield  {author} {\bibinfo {author} {\bibfnamefont {D.}~\bibnamefont
  {Qu}}, \bibinfo {author} {\bibfnamefont {S.~Y.}\ \bibnamefont {Huang}},
  \bibinfo {author} {\bibfnamefont {J.}~\bibnamefont {Hu}}, \bibinfo {author}
  {\bibfnamefont {R.}~\bibnamefont {Wu}}, \ and\ \bibinfo {author}
  {\bibfnamefont {C.~L.}\ \bibnamefont {Chien}},\ }\href@noop {} {\bibfield
  {journal} {\bibinfo  {journal} {Physical Review Letters}\ }\textbf {\bibinfo
  {volume} {110}},\ \bibinfo {pages} {067206} (\bibinfo {year}
  {2013})}\BibitemShut {NoStop}%
\bibitem [{\citenamefont {Kehlberger}\ \emph {et~al.}(2014)\citenamefont
  {Kehlberger}, \citenamefont {Jakob}, \citenamefont {Onbasli}, \citenamefont
  {H~Kim}, \citenamefont {Ross},\ and\ \citenamefont
  {Kl{\"a}ui}}]{Kehlberger:2014ha}%
  \BibitemOpen
  \bibfield  {author} {\bibinfo {author} {\bibfnamefont {A.}~\bibnamefont
  {Kehlberger}}, \bibinfo {author} {\bibfnamefont {G.}~\bibnamefont {Jakob}},
  \bibinfo {author} {\bibfnamefont {M.~C.}\ \bibnamefont {Onbasli}}, \bibinfo
  {author} {\bibfnamefont {D.}~\bibnamefont {H~Kim}}, \bibinfo {author}
  {\bibfnamefont {C.~A.}\ \bibnamefont {Ross}}, \ and\ \bibinfo {author}
  {\bibfnamefont {M.}~\bibnamefont {Kl{\"a}ui}},\ }\href@noop {} {\bibfield
  {journal} {\bibinfo  {journal} {Journal of Applied Physics}\ }\textbf
  {\bibinfo {volume} {115}},\ \bibinfo {pages} {17C731} (\bibinfo {year}
  {2014})}\BibitemShut {NoStop}%
\bibitem [{\citenamefont {Siegel}\ \emph {et~al.}(2014)\citenamefont {Siegel},
  \citenamefont {Prestgard}, \citenamefont {Teng},\ and\ \citenamefont
  {Tiwari}}]{Siegel:2014dx}%
  \BibitemOpen
  \bibfield  {author} {\bibinfo {author} {\bibfnamefont {G.}~\bibnamefont
  {Siegel}}, \bibinfo {author} {\bibfnamefont {M.~C.}\ \bibnamefont
  {Prestgard}}, \bibinfo {author} {\bibfnamefont {S.}~\bibnamefont {Teng}}, \
  and\ \bibinfo {author} {\bibfnamefont {A.}~\bibnamefont {Tiwari}},\
  }\href@noop {} {\bibfield  {journal} {\bibinfo  {journal} {Scientific
  Reports}\ }\textbf {\bibinfo {volume} {4}} (\bibinfo {year}
  {2014})}\BibitemShut {NoStop}%
\bibitem [{\citenamefont {Agrawal}\ \emph {et~al.}(2014)\citenamefont
  {Agrawal}, \citenamefont {Vasyuchka}, \citenamefont {Serga}, \citenamefont
  {Kirihara}, \citenamefont {Pirro}, \citenamefont {Langner}, \citenamefont
  {Jungfleisch}, \citenamefont {Chumak}, \citenamefont {Papaioannou},\ and\
  \citenamefont {Hillebrands}}]{Agrawal:2014ik}%
  \BibitemOpen
  \bibfield  {author} {\bibinfo {author} {\bibfnamefont {M.}~\bibnamefont
  {Agrawal}}, \bibinfo {author} {\bibfnamefont {V.~I.}\ \bibnamefont
  {Vasyuchka}}, \bibinfo {author} {\bibfnamefont {A.~A.}\ \bibnamefont
  {Serga}}, \bibinfo {author} {\bibfnamefont {A.}~\bibnamefont {Kirihara}},
  \bibinfo {author} {\bibfnamefont {P.}~\bibnamefont {Pirro}}, \bibinfo
  {author} {\bibfnamefont {T.}~\bibnamefont {Langner}}, \bibinfo {author}
  {\bibfnamefont {M.~B.}\ \bibnamefont {Jungfleisch}}, \bibinfo {author}
  {\bibfnamefont {A.~V.}\ \bibnamefont {Chumak}}, \bibinfo {author}
  {\bibfnamefont {E.~T.}\ \bibnamefont {Papaioannou}}, \ and\ \bibinfo {author}
  {\bibfnamefont {B.}~\bibnamefont {Hillebrands}},\ }\href@noop {} {\bibfield
  {journal} {\bibinfo  {journal} {Physical Review B}\ }\textbf {\bibinfo
  {volume} {89}},\ \bibinfo {pages} {224414} (\bibinfo {year}
  {2014})}\BibitemShut {NoStop}%
\bibitem [{\citenamefont {Bosu}\ \emph {et~al.}(2011)\citenamefont {Bosu},
  \citenamefont {Sakuraba}, \citenamefont {Uchida}, \citenamefont {Saito},
  \citenamefont {Ota}, \citenamefont {Saitoh},\ and\ \citenamefont
  {Takanashi}}]{Bosu:2011bw}%
  \BibitemOpen
  \bibfield  {author} {\bibinfo {author} {\bibfnamefont {S.}~\bibnamefont
  {Bosu}}, \bibinfo {author} {\bibfnamefont {Y.}~\bibnamefont {Sakuraba}},
  \bibinfo {author} {\bibfnamefont {K.}~\bibnamefont {Uchida}}, \bibinfo
  {author} {\bibfnamefont {K.}~\bibnamefont {Saito}}, \bibinfo {author}
  {\bibfnamefont {T.}~\bibnamefont {Ota}}, \bibinfo {author} {\bibfnamefont
  {E.}~\bibnamefont {Saitoh}}, \ and\ \bibinfo {author} {\bibfnamefont
  {K.}~\bibnamefont {Takanashi}},\ }\href@noop {} {\bibfield  {journal}
  {\bibinfo  {journal} {Physical Review B}\ }\textbf {\bibinfo {volume} {83}}
  (\bibinfo {year} {2011})}\BibitemShut {NoStop}%
\bibitem [{\citenamefont {Huang}\ \emph {et~al.}(2011)\citenamefont {Huang},
  \citenamefont {Wang}, \citenamefont {Lee}, \citenamefont {Kwo},\ and\
  \citenamefont {Chien}}]{Huang:2011cd}%
  \BibitemOpen
  \bibfield  {author} {\bibinfo {author} {\bibfnamefont {S.}~\bibnamefont
  {Huang}}, \bibinfo {author} {\bibfnamefont {W.}~\bibnamefont {Wang}},
  \bibinfo {author} {\bibfnamefont {S.}~\bibnamefont {Lee}}, \bibinfo {author}
  {\bibfnamefont {J.}~\bibnamefont {Kwo}}, \ and\ \bibinfo {author}
  {\bibfnamefont {C.}~\bibnamefont {Chien}},\ }\href@noop {} {\bibfield
  {journal} {\bibinfo  {journal} {Physical Review Letters}\ }\textbf {\bibinfo
  {volume} {107}},\ \bibinfo {pages} {196602} (\bibinfo {year}
  {2011})}\BibitemShut {NoStop}%
\bibitem [{\citenamefont {Schmid}\ \emph {et~al.}(2013)\citenamefont {Schmid},
  \citenamefont {Srichandan}, \citenamefont {Meier}, \citenamefont {Kuschel},
  \citenamefont {Schmalhorst}, \citenamefont {Vogel}, \citenamefont {Reiss},
  \citenamefont {Strunk},\ and\ \citenamefont {Back}}]{Schmid:2013m}%
  \BibitemOpen
  \bibfield  {author} {\bibinfo {author} {\bibfnamefont {M.}~\bibnamefont
  {Schmid}}, \bibinfo {author} {\bibfnamefont {S.}~\bibnamefont {Srichandan}},
  \bibinfo {author} {\bibfnamefont {D.}~\bibnamefont {Meier}}, \bibinfo
  {author} {\bibfnamefont {T.}~\bibnamefont {Kuschel}}, \bibinfo {author}
  {\bibfnamefont {J.~M.}\ \bibnamefont {Schmalhorst}}, \bibinfo {author}
  {\bibfnamefont {M.}~\bibnamefont {Vogel}}, \bibinfo {author} {\bibfnamefont
  {G.}~\bibnamefont {Reiss}}, \bibinfo {author} {\bibfnamefont
  {C.}~\bibnamefont {Strunk}}, \ and\ \bibinfo {author} {\bibfnamefont {C.~H.}\
  \bibnamefont {Back}},\ }\href@noop {} {\bibfield  {journal} {\bibinfo
  {journal} {Physical Review Letters}\ }\textbf {\bibinfo {volume} {111}},\
  \bibinfo {pages} {187201} (\bibinfo {year} {2013})}\BibitemShut {NoStop}%
\bibitem [{\citenamefont {Bui}\ and\ \citenamefont
  {Rivadulla}(2014)}]{Bui:2014ab}%
  \BibitemOpen
  \bibfield  {author} {\bibinfo {author} {\bibfnamefont {C.~T.}\ \bibnamefont
  {Bui}}\ and\ \bibinfo {author} {\bibfnamefont {F.}~\bibnamefont
  {Rivadulla}},\ }\href {\doibase 10.1103/PhysRevB.90.100403} {\bibfield
  {journal} {\bibinfo  {journal} {Phys. Rev. B}\ }\textbf {\bibinfo {volume}
  {90}},\ \bibinfo {pages} {100403} (\bibinfo {year} {2014})}\BibitemShut
  {NoStop}%
\bibitem [{\citenamefont {Soldatov}\ \emph {et~al.}(2014)\citenamefont
  {Soldatov}, \citenamefont {Panarina}, \citenamefont {Hess}, \citenamefont
  {Schultz},\ and\ \citenamefont {Sch\"afer}}]{Soldatov:2014ab}%
  \BibitemOpen
  \bibfield  {author} {\bibinfo {author} {\bibfnamefont {I.~V.}\ \bibnamefont
  {Soldatov}}, \bibinfo {author} {\bibfnamefont {N.}~\bibnamefont {Panarina}},
  \bibinfo {author} {\bibfnamefont {C.}~\bibnamefont {Hess}}, \bibinfo {author}
  {\bibfnamefont {L.}~\bibnamefont {Schultz}}, \ and\ \bibinfo {author}
  {\bibfnamefont {R.}~\bibnamefont {Sch\"afer}},\ }\href {\doibase
  10.1103/PhysRevB.90.104423} {\bibfield  {journal} {\bibinfo  {journal} {Phys.
  Rev. B}\ }\textbf {\bibinfo {volume} {90}},\ \bibinfo {pages} {104423}
  (\bibinfo {year} {2014})}\BibitemShut {NoStop}%
\bibitem [{\citenamefont {Schreier}\ \emph {et~al.}(2013)\citenamefont
  {Schreier}, \citenamefont {Kamra}, \citenamefont {Weiler}, \citenamefont
  {Xiao}, \citenamefont {Bauer}, \citenamefont {Gross},\ and\ \citenamefont
  {Goennenwein}}]{Schreier:2013ab}%
  \BibitemOpen
  \bibfield  {author} {\bibinfo {author} {\bibfnamefont {M.}~\bibnamefont
  {Schreier}}, \bibinfo {author} {\bibfnamefont {A.}~\bibnamefont {Kamra}},
  \bibinfo {author} {\bibfnamefont {M.}~\bibnamefont {Weiler}}, \bibinfo
  {author} {\bibfnamefont {J.}~\bibnamefont {Xiao}}, \bibinfo {author}
  {\bibfnamefont {G.~E.~W.}\ \bibnamefont {Bauer}}, \bibinfo {author}
  {\bibfnamefont {R.}~\bibnamefont {Gross}}, \ and\ \bibinfo {author}
  {\bibfnamefont {S.~T.~B.}\ \bibnamefont {Goennenwein}},\ }\href {\doibase
  10.1103/PhysRevB.88.094410} {\bibfield  {journal} {\bibinfo  {journal} {Phys.
  Rev. B}\ }\textbf {\bibinfo {volume} {88}},\ \bibinfo {pages} {094410}
  (\bibinfo {year} {2013})}\BibitemShut {NoStop}%
\bibitem [{\citenamefont {Li}\ \emph {et~al.}(2011)\citenamefont {Li},
  \citenamefont {Wang}, \citenamefont {Iliev}, \citenamefont {Klein},\ and\
  \citenamefont {Gupta}}]{Li:2011ka}%
  \BibitemOpen
  \bibfield  {author} {\bibinfo {author} {\bibfnamefont {N.}~\bibnamefont
  {Li}}, \bibinfo {author} {\bibfnamefont {Y.-H.~A.}\ \bibnamefont {Wang}},
  \bibinfo {author} {\bibfnamefont {M.~N.}\ \bibnamefont {Iliev}}, \bibinfo
  {author} {\bibfnamefont {T.~M.}\ \bibnamefont {Klein}}, \ and\ \bibinfo
  {author} {\bibfnamefont {A.}~\bibnamefont {Gupta}},\ }\href@noop {}
  {\bibfield  {journal} {\bibinfo  {journal} {Chemical Vapor Deposition}\
  }\textbf {\bibinfo {volume} {17}},\ \bibinfo {pages} {261} (\bibinfo {year}
  {2011})}\BibitemShut {NoStop}%
\bibitem [{\citenamefont {Schreier}\ \emph {et~al.}(2014)\citenamefont
  {Schreier}, \citenamefont {Bauer}, \citenamefont {Vasyuchka}, \citenamefont
  {Flipse}, \citenamefont {Uchida}, \citenamefont {Lotze}, \citenamefont
  {Lauer}, \citenamefont {Chumak}, \citenamefont {Serga}, \citenamefont
  {Daimon}, \citenamefont {Kikkawa}, \citenamefont {Saitoh}, \citenamefont {van
  Wees}, \citenamefont {Hillebrands}, \citenamefont {Gross},\ and\
  \citenamefont {Goennenwein}}]{Schreier:2014ax}%
  \BibitemOpen
  \bibfield  {author} {\bibinfo {author} {\bibfnamefont {M.}~\bibnamefont
  {Schreier}}, \bibinfo {author} {\bibfnamefont {G.~E.~W.}\ \bibnamefont
  {Bauer}}, \bibinfo {author} {\bibfnamefont {V.}~\bibnamefont {Vasyuchka}},
  \bibinfo {author} {\bibfnamefont {J.}~\bibnamefont {Flipse}}, \bibinfo
  {author} {\bibfnamefont {K.}~\bibnamefont {Uchida}}, \bibinfo {author}
  {\bibfnamefont {J.}~\bibnamefont {Lotze}}, \bibinfo {author} {\bibfnamefont
  {V.}~\bibnamefont {Lauer}}, \bibinfo {author} {\bibfnamefont
  {A.}~\bibnamefont {Chumak}}, \bibinfo {author} {\bibfnamefont
  {A.}~\bibnamefont {Serga}}, \bibinfo {author} {\bibfnamefont
  {S.}~\bibnamefont {Daimon}}, \bibinfo {author} {\bibfnamefont
  {T.}~\bibnamefont {Kikkawa}}, \bibinfo {author} {\bibfnamefont
  {E.}~\bibnamefont {Saitoh}}, \bibinfo {author} {\bibfnamefont {B.~J.}\
  \bibnamefont {van Wees}}, \bibinfo {author} {\bibfnamefont {B.}~\bibnamefont
  {Hillebrands}}, \bibinfo {author} {\bibfnamefont {R.}~\bibnamefont {Gross}},
  \ and\ \bibinfo {author} {\bibfnamefont {S.~T.~B.}\ \bibnamefont
  {Goennenwein}},\ }\href@noop {} {\bibfield  {journal} {\bibinfo  {journal}
  {arXiv:1404.3490}\ } (\bibinfo {year} {2014})}\BibitemShut {NoStop}%
\bibitem [{\citenamefont {Xiao}\ \emph {et~al.}(2010)\citenamefont {Xiao},
  \citenamefont {Bauer}, \citenamefont {Uchida}, \citenamefont {Saitoh},\ and\
  \citenamefont {Maekawa}}]{Xiao:2010iy}%
  \BibitemOpen
  \bibfield  {author} {\bibinfo {author} {\bibfnamefont {J.}~\bibnamefont
  {Xiao}}, \bibinfo {author} {\bibfnamefont {G.}~\bibnamefont {Bauer}},
  \bibinfo {author} {\bibfnamefont {K.}~\bibnamefont {Uchida}}, \bibinfo
  {author} {\bibfnamefont {E.}~\bibnamefont {Saitoh}}, \ and\ \bibinfo {author}
  {\bibfnamefont {S.}~\bibnamefont {Maekawa}},\ }\href@noop {} {\bibfield
  {journal} {\bibinfo  {journal} {Physical Review B}\ }\textbf {\bibinfo
  {volume} {81}} (\bibinfo {year} {2010})}\BibitemShut {NoStop}%
\bibitem [{\citenamefont {Flipse}\ \emph {et~al.}(2012)\citenamefont {Flipse},
  \citenamefont {Bakker}, \citenamefont {Slachter}, \citenamefont {Dejene},\
  and\ \citenamefont {van Wees}}]{Flipse:2012kn}%
  \BibitemOpen
  \bibfield  {author} {\bibinfo {author} {\bibfnamefont {J.}~\bibnamefont
  {Flipse}}, \bibinfo {author} {\bibfnamefont {F.~L.}\ \bibnamefont {Bakker}},
  \bibinfo {author} {\bibfnamefont {A.}~\bibnamefont {Slachter}}, \bibinfo
  {author} {\bibfnamefont {F.~K.}\ \bibnamefont {Dejene}}, \ and\ \bibinfo
  {author} {\bibfnamefont {B.~J.}\ \bibnamefont {van Wees}},\ }\href@noop {}
  {\bibfield  {journal} {\bibinfo  {journal} {Nature Nanotechnology}\ }\textbf
  {\bibinfo {volume} {7}},\ \bibinfo {pages} {166} (\bibinfo {year}
  {2012})}\BibitemShut {NoStop}%
\bibitem [{\citenamefont {Weiler}\ \emph {et~al.}(2013)\citenamefont {Weiler},
  \citenamefont {Althammer}, \citenamefont {Schreier}, \citenamefont {Lotze},
  \citenamefont {Pernpeintner}, \citenamefont {Meyer}, \citenamefont {Huebl},
  \citenamefont {Gross}, \citenamefont {Kamra}, \citenamefont {Xiao},
  \citenamefont {Chen}, \citenamefont {Jiao}, \citenamefont {Bauer},\ and\
  \citenamefont {Goennenwein}}]{Weiler:2013ko}%
  \BibitemOpen
  \bibfield  {author} {\bibinfo {author} {\bibfnamefont {M.}~\bibnamefont
  {Weiler}}, \bibinfo {author} {\bibfnamefont {M.}~\bibnamefont {Althammer}},
  \bibinfo {author} {\bibfnamefont {M.}~\bibnamefont {Schreier}}, \bibinfo
  {author} {\bibfnamefont {J.}~\bibnamefont {Lotze}}, \bibinfo {author}
  {\bibfnamefont {M.}~\bibnamefont {Pernpeintner}}, \bibinfo {author}
  {\bibfnamefont {S.}~\bibnamefont {Meyer}}, \bibinfo {author} {\bibfnamefont
  {H.}~\bibnamefont {Huebl}}, \bibinfo {author} {\bibfnamefont
  {R.}~\bibnamefont {Gross}}, \bibinfo {author} {\bibfnamefont
  {A.}~\bibnamefont {Kamra}}, \bibinfo {author} {\bibfnamefont
  {J.}~\bibnamefont {Xiao}}, \bibinfo {author} {\bibfnamefont {Y.-T.}\
  \bibnamefont {Chen}}, \bibinfo {author} {\bibfnamefont {H.}~\bibnamefont
  {Jiao}}, \bibinfo {author} {\bibfnamefont {G.~E.~W.}\ \bibnamefont {Bauer}},
  \ and\ \bibinfo {author} {\bibfnamefont {S.~T.~B.}\ \bibnamefont
  {Goennenwein}},\ }\href@noop {} {\bibfield  {journal} {\bibinfo  {journal}
  {Physical Review Letters}\ }\textbf {\bibinfo {volume} {111}},\ \bibinfo
  {pages} {176601} (\bibinfo {year} {2013})}\BibitemShut {NoStop}%
\bibitem [{\citenamefont {Qiu}\ \emph {et~al.}(2013)\citenamefont {Qiu},
  \citenamefont {Ando}, \citenamefont {Uchida}, \citenamefont {Kajiwara},
  \citenamefont {Takahashi}, \citenamefont {Nakayama}, \citenamefont {An},
  \citenamefont {Fujikawa},\ and\ \citenamefont {Saitoh}}]{Qiu:2013je}%
  \BibitemOpen
  \bibfield  {author} {\bibinfo {author} {\bibfnamefont {Z.}~\bibnamefont
  {Qiu}}, \bibinfo {author} {\bibfnamefont {K.}~\bibnamefont {Ando}}, \bibinfo
  {author} {\bibfnamefont {K.}~\bibnamefont {Uchida}}, \bibinfo {author}
  {\bibfnamefont {Y.}~\bibnamefont {Kajiwara}}, \bibinfo {author}
  {\bibfnamefont {R.}~\bibnamefont {Takahashi}}, \bibinfo {author}
  {\bibfnamefont {H.}~\bibnamefont {Nakayama}}, \bibinfo {author}
  {\bibfnamefont {T.}~\bibnamefont {An}}, \bibinfo {author} {\bibfnamefont
  {Y.}~\bibnamefont {Fujikawa}}, \ and\ \bibinfo {author} {\bibfnamefont
  {E.}~\bibnamefont {Saitoh}},\ }\href@noop {} {\bibfield  {journal} {\bibinfo
  {journal} {Applied Physics Letters}\ }\textbf {\bibinfo {volume} {103}},\
  \bibinfo {pages} {092404} (\bibinfo {year} {2013})}\BibitemShut {NoStop}%
\bibitem [{\citenamefont {Klewe}\ \emph {et~al.}(2014)\citenamefont {Klewe},
  \citenamefont {Meinert}, \citenamefont {Boehnke}, \citenamefont {Kuepper},
  \citenamefont {Arenholz}, \citenamefont {Gupta}, \citenamefont {Schmalhorst},
  \citenamefont {Kuschel},\ and\ \citenamefont {Reiss}}]{Klewe:2014km}%
  \BibitemOpen
  \bibfield  {author} {\bibinfo {author} {\bibfnamefont {C.}~\bibnamefont
  {Klewe}}, \bibinfo {author} {\bibfnamefont {M.}~\bibnamefont {Meinert}},
  \bibinfo {author} {\bibfnamefont {A.}~\bibnamefont {Boehnke}}, \bibinfo
  {author} {\bibfnamefont {K.}~\bibnamefont {Kuepper}}, \bibinfo {author}
  {\bibfnamefont {E.}~\bibnamefont {Arenholz}}, \bibinfo {author}
  {\bibfnamefont {A.}~\bibnamefont {Gupta}}, \bibinfo {author} {\bibfnamefont
  {J.~M.}\ \bibnamefont {Schmalhorst}}, \bibinfo {author} {\bibfnamefont
  {T.}~\bibnamefont {Kuschel}}, \ and\ \bibinfo {author} {\bibfnamefont
  {G.}~\bibnamefont {Reiss}},\ }\href@noop {} {\bibfield  {journal} {\bibinfo
  {journal} {Journal of Applied Physics}\ }\textbf {\bibinfo {volume} {115}},\
  \bibinfo {pages} {123903} (\bibinfo {year} {2014})}\BibitemShut {NoStop}%
\bibitem [{\citenamefont {Kuschel}\ \emph {et~al.}(2014)\citenamefont
  {Kuschel}, \citenamefont {Klewe}, \citenamefont {Schmalhorst}, \citenamefont
  {Bertram}, \citenamefont {Schuckmann}, \citenamefont {Schemme}, \citenamefont
  {Wollschl{\"a}ger}, \citenamefont {Francoual}, \citenamefont {Strempfer},
  \citenamefont {Gupta}, \citenamefont {Meinert}, \citenamefont {G{\"o}tz},
  \citenamefont {Meier},\ and\ \citenamefont {Reiss}}]{Kuschel:2014ab}%
  \BibitemOpen
  \bibfield  {author} {\bibinfo {author} {\bibfnamefont {T.}~\bibnamefont
  {Kuschel}}, \bibinfo {author} {\bibfnamefont {C.}~\bibnamefont {Klewe}},
  \bibinfo {author} {\bibfnamefont {J.~M.}\ \bibnamefont {Schmalhorst}},
  \bibinfo {author} {\bibfnamefont {F.}~\bibnamefont {Bertram}}, \bibinfo
  {author} {\bibfnamefont {O.}~\bibnamefont {Schuckmann}}, \bibinfo {author}
  {\bibfnamefont {T.}~\bibnamefont {Schemme}}, \bibinfo {author} {\bibfnamefont
  {J.}~\bibnamefont {Wollschl{\"a}ger}}, \bibinfo {author} {\bibfnamefont
  {S.}~\bibnamefont {Francoual}}, \bibinfo {author} {\bibfnamefont
  {J.}~\bibnamefont {Strempfer}}, \bibinfo {author} {\bibfnamefont
  {A.}~\bibnamefont {Gupta}}, \bibinfo {author} {\bibfnamefont
  {M.}~\bibnamefont {Meinert}}, \bibinfo {author} {\bibfnamefont
  {G.}~\bibnamefont {G{\"o}tz}}, \bibinfo {author} {\bibfnamefont
  {D.}~\bibnamefont {Meier}}, \ and\ \bibinfo {author} {\bibfnamefont
  {G.}~\bibnamefont {Reiss}},\ }\href@noop {} {\bibfield  {journal} {\bibinfo
  {journal} {arXiv:1411.0113}\ } (\bibinfo {year} {2014})}\BibitemShut
  {NoStop}%
\bibitem [{\citenamefont {Gepr{\"a}gs}\ \emph {et~al.}(2012)\citenamefont
  {Gepr{\"a}gs}, \citenamefont {Meyer}, \citenamefont {Altmannshofer},
  \citenamefont {Opel}, \citenamefont {Wilhelm}, \citenamefont {Rogalev},
  \citenamefont {Gross},\ and\ \citenamefont {Goennenwein}}]{Geprags:2012be}%
  \BibitemOpen
  \bibfield  {author} {\bibinfo {author} {\bibfnamefont {S.}~\bibnamefont
  {Gepr{\"a}gs}}, \bibinfo {author} {\bibfnamefont {S.}~\bibnamefont {Meyer}},
  \bibinfo {author} {\bibfnamefont {S.}~\bibnamefont {Altmannshofer}}, \bibinfo
  {author} {\bibfnamefont {M.}~\bibnamefont {Opel}}, \bibinfo {author}
  {\bibfnamefont {F.}~\bibnamefont {Wilhelm}}, \bibinfo {author} {\bibfnamefont
  {A.}~\bibnamefont {Rogalev}}, \bibinfo {author} {\bibfnamefont
  {R.}~\bibnamefont {Gross}}, \ and\ \bibinfo {author} {\bibfnamefont
  {S.~T.~B.}\ \bibnamefont {Goennenwein}},\ }\href@noop {} {\bibfield
  {journal} {\bibinfo  {journal} {Applied Physics Letters}\ }\textbf {\bibinfo
  {volume} {101}},\ \bibinfo {pages} {262407} (\bibinfo {year}
  {2012})}\BibitemShut {NoStop}%
\bibitem [{\citenamefont {Lu}\ \emph {et~al.}(2013)\citenamefont {Lu},
  \citenamefont {Choi}, \citenamefont {Ortega}, \citenamefont {Cheng},
  \citenamefont {Cai}, \citenamefont {Huang}, \citenamefont {Sun},\ and\
  \citenamefont {Chien}}]{Lu:2013eq}%
  \BibitemOpen
  \bibfield  {author} {\bibinfo {author} {\bibfnamefont {Y.}~\bibnamefont
  {Lu}}, \bibinfo {author} {\bibfnamefont {Y.}~\bibnamefont {Choi}}, \bibinfo
  {author} {\bibfnamefont {C.}~\bibnamefont {Ortega}}, \bibinfo {author}
  {\bibfnamefont {X.}~\bibnamefont {Cheng}}, \bibinfo {author} {\bibfnamefont
  {J.}~\bibnamefont {Cai}}, \bibinfo {author} {\bibfnamefont {S.}~\bibnamefont
  {Huang}}, \bibinfo {author} {\bibfnamefont {L.}~\bibnamefont {Sun}}, \ and\
  \bibinfo {author} {\bibfnamefont {C.}~\bibnamefont {Chien}},\ }\href@noop {}
  {\bibfield  {journal} {\bibinfo  {journal} {Physical Review Letters}\
  }\textbf {\bibinfo {volume} {110}},\ \bibinfo {pages} {147207} (\bibinfo
  {year} {2013})}\BibitemShut {NoStop}%
\bibitem [{\citenamefont {Kikkawa}\ \emph {et~al.}(2013)\citenamefont
  {Kikkawa}, \citenamefont {Uchida}, \citenamefont {Shiomi}, \citenamefont
  {Qiu}, \citenamefont {Hou}, \citenamefont {Tian}, \citenamefont {Nakayama},
  \citenamefont {Jin},\ and\ \citenamefont {Saitoh}}]{Kikkawa:2013hx}%
  \BibitemOpen
  \bibfield  {author} {\bibinfo {author} {\bibfnamefont {T.}~\bibnamefont
  {Kikkawa}}, \bibinfo {author} {\bibfnamefont {K.}~\bibnamefont {Uchida}},
  \bibinfo {author} {\bibfnamefont {Y.}~\bibnamefont {Shiomi}}, \bibinfo
  {author} {\bibfnamefont {Z.}~\bibnamefont {Qiu}}, \bibinfo {author}
  {\bibfnamefont {D.}~\bibnamefont {Hou}}, \bibinfo {author} {\bibfnamefont
  {D.}~\bibnamefont {Tian}}, \bibinfo {author} {\bibfnamefont {H.}~\bibnamefont
  {Nakayama}}, \bibinfo {author} {\bibfnamefont {X.~F.}\ \bibnamefont {Jin}}, \
  and\ \bibinfo {author} {\bibfnamefont {E.}~\bibnamefont {Saitoh}},\
  }\href@noop {} {\bibfield  {journal} {\bibinfo  {journal} {Physical Review
  Letters}\ }\textbf {\bibinfo {volume} {110}},\ \bibinfo {pages} {067207}
  (\bibinfo {year} {2013})}\BibitemShut {NoStop}%
\bibitem [{\citenamefont {Wegrowe}\ \emph {et~al.}(2014)\citenamefont
  {Wegrowe}, \citenamefont {Drouhin},\ and\ \citenamefont
  {Lacour}}]{Wegrowe:2014bs}%
  \BibitemOpen
  \bibfield  {author} {\bibinfo {author} {\bibfnamefont {J.~E.}\ \bibnamefont
  {Wegrowe}}, \bibinfo {author} {\bibfnamefont {H.~J.}\ \bibnamefont
  {Drouhin}}, \ and\ \bibinfo {author} {\bibfnamefont {D.}~\bibnamefont
  {Lacour}},\ }\href@noop {} {\bibfield  {journal} {\bibinfo  {journal}
  {Physical Review B}\ }\textbf {\bibinfo {volume} {89}},\ \bibinfo {pages}
  {094409} (\bibinfo {year} {2014})}\BibitemShut {NoStop}%
\end{thebibliography}%

\end{document}